\documentstyle[11pt]{article}

\newlength{\dinwidth}
\newlength{\dinmargin}
\def\slepton{\widetilde \ell}
\def\sl{{\widetilde \ell}^{-}}

\def\sneu{\widetilde \nu}

\def\st{\widetilde{t}}
\def\sb{\widetilde{b}}

\def\sz1{{\widetilde{Z}}_{1}}

\def\szk{{\widetilde{Z}}_{k}}

\def\swi{{\widetilde{W}}_{i}}

\def\msz1{m_{\sz1}}

\def\nle{{\stackrel{<}{\sim}}}
\def\nge{{\stackrel{>}{\sim}}}

\def\stl{\st_{1}}

\def\sth{\st_{2}}

\def\mstl{m_{\stl}}
\def\msth{m_{\sth}}

\def\tht{\theta_{t}}

\def\rb{\slash\hspace{-7pt}R}

\begin{document}
~~~\\
\vspace{10mm}
\begin{flushright}
ITP-SU-97/02  \\
hep-ph/9704221 \\
April, 1997
\end{flushright}
\begin{center}
  \begin{Large}
   \begin{bf}
Stops in R-parity Breaking Model for High-$Q^2$ Events at HERA \\
   \end{bf}
  \end{Large}
  \vspace{5mm}
  \begin{large}
Tadashi Kon \\
  \end{large}
Faculty of Engineering, Seikei University, Tokyo 180, Japan \\
kon@ge.seikei.ac.jp\\
 \vspace{3mm}
 \begin{large}
    Tetsuro Kobayashi\\
  \end{large}
Faculty of Engineering, 
Fukui Institute of Technology, Fukui 910, Japan \\
koba@ge.seikei.ac.jp\\
  \vspace{5mm}
\end{center}
\vskip50pt
\begin{quotation}
\noindent
\begin{center}
{\bf Abstract}
\end{center}
We investigate an event excess in the reaction $e^+p \to e^+ X$ with large $x$ 
and high $Q^2$ observed 
at HERA and show that the events could naturally be interpreted as a signature 
of the 
production of the scalar top quarks (stops) in a supersymmetric model with 
$R$-parity breaking interactions. The HERA events are characterized by the 
broad 
mass distribution and in fact it can be simulated by our specific scenario if 
we 
consider almost degenerate two mass eigenstates $\widetilde{t_1}$ and 
$\widetilde{t_2}$ 
of the stops. 
\end{quotation}
\vfill\eject
%
%

Recently both the H1 and ZEUS Collaboration reported that an event 
excess in the deep inelastic scattering (DIS) 
$e^+p \to e^+ X$ with large $x$ and high $Q^2$ had 
been observed \cite{H1,ZEUS} at HERA. 
The total data sample analyzed at the H1 and ZEUS corresponds to respectively 
accumulated luminosity of 14.2pb$^{-1}$ and 20.1pb$^{-1}$ 
in positron (27.5GeV) -- proton (820GeV) collisions. 
For high $Q^2$ ($>$ $15,000$GeV$^2$) region 12 events are observed at both 
the H1 and ZEUS, while the Standard Model expectation are 4.7 $\pm$ 0.8 and 
8.7 $\pm$ 0.7, respectively. 
Combining these results we get $N_{obs}$ $=$ 24 ($\sigma_{obs}$ $\simeq$ 
0.71pb) against $N_{exp}$ $=$ 13 $\pm$ 1 ($\sigma_{exp}$ $\simeq$ 0.38pb). 

The event excess is characterized by not only a high four momemtum transfer 
$Q^2$ but also a large Bjorken parameter $x$ $\nge$ 0.3. 
Interestingly, the events seem to cluster at invariant mass $M$ $=$ 200 $\sim$ 
220GeV, 
where $M$ is directly determined from $x$ by 
$M$ $\equiv$ $\sqrt{xs}$. 
However, while the excess events of the H1 seem to cluster at 
$M$ $\simeq$ 200GeV, those of the ZEUS have a broader distribution around 
$M$ $\simeq$ 220GeV. 
Statistical consistency between the H1 and ZEUS data was discussed in 
ref.\cite{drees}, where a possible single resonance structure was assumed. 
Here we will simply combine both data since each statistical significance 
is still not so high enough.

Some possible interpretations of the events have already been given ; 
\begin{description}
\item[(A)] four fermion contact interaction \cite{alt,cnt,sqlqcnt}
\item[(B)] leptoquark production  \cite{drees,alt,kali,lq,sqlqcnt,sqlq}
\item[(C)] scalar quark production in the supersymmetric (SUSY) model with 
R-parity 
breaking (RB) interactions \cite{chou,drm,alt,kali,sq,sqlqcnt,sqlq}
\end {description}
In this letter we investigate one of scenarios of the category (C), 
the scalar top (stop) \cite{stop} production in the RB SUSY model 
\cite{chou,drm,alt,kali}. 
For we have first proposed the scenario at HERA experiments long since 
\cite{stoprb,stopsg}. 
On the other hand, 
possibilities of the search for the squarks in the first or second generation 
at HERA have originaly been discussed in ref.\cite{hewett,dreiner}. 


The discussion is based on the minimal SUSY standard model 
(MSSM) with an RB interaction  
\begin{equation}
L=\lambda'_{1jk} ({\widetilde{u_{jL}}} {\overline{d_k}} P_L e 
- {\overline{\widetilde{d_{kR}}}} \overline{e^c} P_L u_j) + h.c,  
\label{sqRb}
\end{equation}
where $P_{L,R}$ read left and right  handed chiral projection operators, 
respectively.   
The interaction Lagrangian (\ref{sqRb}) has been derived from the general RB 
superpotential  \cite{Barger}; 
\begin{equation}
W_{\rb}=\lambda_{ijk}\hat{L}_i \hat{L}_j \hat{E^c}_k 
+ \lambda'_{ijk}\hat{L}_i \hat{Q}_j \hat{D^c}_k + 
\lambda''_{ijk}\hat{U^c}_i \hat{D^c}_j \hat{D^c}_k, 
\label{RBW}
\end{equation}
where $i, j, k$ are generation indices. 
The first two terms violate the lepton number $L$ and the last term 
violates the baryon number $B$. 
If we want to explain such unresolved problems as 
({\romannumeral 1}) the cosmic baryon number violation, 
({\romannumeral 2}) the origin of the masses and the 
magnetic moments of neutrinos and 
({\romannumeral 3}) some interesting rare processes 
in terms of the $L$ and/or $B$ violation, 
$R$-parity breaking terms must be incorporated in the MSSM. 
The coupling Eq. (\ref{sqRb}) 
will be most suitable for the $ep$ collider experiments at HERA 
because the squark ${\widetilde{u_{jL}}}$ or ${\overline{\widetilde{d_{kR}}}}$ 
will be produced in the $s$-channel 
in $e^+$-$q$ sub-processes. 
\begin{eqnarray}
&& e^+ + d_k \to {\widetilde{u_{jL}}} \to e^+ + d_k \\
&& e^+ + {\overline{u_j}} \to {\overline{\widetilde{d_{kR}}}} \to e^+ + 
{\overline{u_j}} .
\end{eqnarray}
The upper bounds on the coupling constants $\lambda'_{1jk}$ have already been 
settled by some experiments, neutrinoless double beta decay \cite{beta}, 
charged current universality \cite{Barger,ccu}, atomic parity violation (APV) 
\cite{Barger,ccu,apv} and $\nu_e$ mass \cite{nemass}. 
Among nine candidate coupling constants $\lambda'_{1jk}$ some are severely 
constrained by these experiments. 
To explain the observed cross section 
($\sim$ 0.7pb) as well as to circumvent the experimental constraints, 
we should consider the two kinds of processes (3) in which a valence 
quark in the proton participates, 
\begin{eqnarray}
&& e^+ + d \to {\widetilde{c_{L}}} \to e^+ + d \\
&& e^+ + d \to {\widetilde{t_{L}}} \to e^+ + d .
\end{eqnarray}
In fact, the stringest upper bounds on 
the corresponding coupling constants coming from the APV experiments are not 
so severe, 
\begin{equation}
\lambda'_{121}, \lambda'_{131} < 0.13
\end{equation}
for $m_{\widetilde{c_L}}$, $m_{\widetilde{t_L}}$ $\simeq$ 200GeV. 
Note that the scalar charm (scharm) ${\widetilde{c_L}}$ and the stop 
${\widetilde{t_L}}$ 
cannot couple to any neutrinos via $R$-breaking interactions. 
That is, no event excesses are expected in the charged current process 
$ep \to \nu q X$ in the squark scenarios. 
This is a unique property of the squark scenarios which could be useful 
for us to distinguish the squarks from leptoquarks. 

Here we pay attention to a fact that the stops 
(${\widetilde{t_L}}$, ${\widetilde{t_R}}$) are naturally mixed each 
other \cite{stop} due to a large top quark mass \cite{top} 
and the mass eigenstates ($\stl$, $\sth$) 
are parametrized by a mixing angle $\tht$, 
\begin{equation}
{\widetilde{t_L}} = \stl\cos\tht - \sth\sin\tht. 
\end{equation}
In this case the interaction Lagrangian (\ref{sqRb}) for the stops is written 
by 
\begin{equation}
L=\lambda'_{131} (\cos\tht \stl {\overline{d}} P_L e 
                - \sin\tht \sth {\overline{d}} P_L e ) + h.c. . 
\label{stRb}
\end{equation}
It should be emphasized that only one scharm ${\widetilde{c_L}}$ could be 
coupled to $e^+ d$ in the scenario with $\lambda'_{121}$ $\neq$ 0. 

Before discussing numerical results, we examine the decay modes of the stop. 
In the MSSM, the stop lighter than the other squarks 
in the 1st and 2nd generations and the gluino 
can decay into various final states : 
\begin{eqnarray*}
\st &\to& t\,\szk   \qquad\qquad\qquad\qquad\qquad\qquad({\rm a}) \\
 &\to& b\,\swi   \ \ \quad\qquad\qquad\qquad\qquad\qquad({\rm b})\\
 &\to& W\,\sb   \ \ \quad\qquad\qquad\qquad\qquad\qquad({\rm c})\\
 &\to& b\,\ell\,\sneu \qquad\qquad\qquad\qquad\qquad\qquad({\rm d})\\
 &\to& b\,\nu\,\slepton \qquad\qquad\qquad\qquad\qquad\qquad({\rm e})\\
 &\to& b\,W\,\szk \ \qquad\qquad\qquad\qquad\qquad \ ({\rm f})\\
 &\to& b\,f\,\overline{f}\,\szk \qquad\qquad\qquad\qquad\qquad
 \ ({\rm g})\\
 &\to& c\,\sz1 \quad\qquad\qquad\qquad\qquad\qquad \ \ ({\rm h})\\
 &\to& e\,d, \ \ \quad\qquad\qquad\qquad\qquad\qquad \ ({\rm i})
\end{eqnarray*}
where $\szk$ ($k=1\sim 4$), $\swi$($i=1,2$), $\sneu$ and $\slepton$, 
respectively, 
denote 
the neutralino, the chargino, the sneutrino and the charged slepton. 
(a) $\sim$ (h) are the $R$-parity conserving decay modes, while (i) is 
only realized through the RB couplings (\ref{stRb}). 

If we consider the RB coupling with $\lambda'_{131}$ $>$ $0.01$, 
the decay modes (d) to (h) are negligible due to their large power of 
$\alpha$ arising from 
multiparticle final state or one loop contribution. 
Moreover, in the present case ($\mstl$ $\simeq$ 200GeV) the mode (a) will be 
kinematically suppressed. 
Then only two body decay modes (b), (c) and (i) are left for our purpose.


In Fig.1 we show the $\lambda'_{131}$ dependence of the total cross section 
for the process, 
\begin{equation}
e^+ p \to \stl X \to e^+ q X, 
\end{equation}
where we take $\mstl$ $=$ 210GeV, $\tht$ $=$ 0 and $Q^2$ $>$ 15,000GeV$^2$. 
The horizontal lines correspond to the observed cross section at HERA, 
$\sigma_{obs}$ $=$ (0.71 $\pm$ 0.15) pb. 
We find the HERA data could be explained by 
$\lambda'_{131}$ $\simeq$ 0.04, 0.06 and 0.13 for 
Br($\stl \to e^+ d$) = 1.0, 0.5 and 0.1, respectively. 
This is almost the same result with that given by Altarelli et al. \cite{alt}. 
From the Fig.1 we see that 0.03 $\nle$ $\lambda'_{131}$ $<$ 0.13 and 
Br($\stl \to e^+ d$) $\nge$ 0.05 are needed for us to obtain the required 
amount of the cross section of about 0.7pb. 
These facts immediately lead us to another important consequence, 
$\Gamma_{tot}(\st)$ $\nle$ 1.5GeV, because 
\begin{equation}
\Gamma(\stl \to e^+ d) = {\frac{(\lambda'_{131}\cos\tht)^2}{16\pi}}\mstl 
             \nle 0.07 \quad {\rm GeV} 
\end{equation}
and $\Gamma_{extra}$ $<$ $(1-0.05)/0.05$ $\Gamma(\stl \to e^+ d)$, 
where $\Gamma_{extra}$ $\equiv$ $\Gamma_{tot}(\st)$ $-$ 
$\Gamma(\stl \to e^+ d)$. 
Note that results obtained here are not altered even if we consider 
the stop mixing $\tht \neq 0$. Moreover, they are applicable to the scharm 
scenario 
when we simply replace ($\stl\to {\widetilde{c_L}}$, 
$\lambda'_{131}\to\lambda'_{121}$).

\renewcommand{\thefootnote}{\fnsymbol{footnote}}
Br($\st \to e^+ d$) = 1.0 can easily be realized for 
$m_{\widetilde{t}}$ $<$ $m_b+m_{\widetilde{W}_1}$ and 
$m_{\widetilde{t}}$ $<$ $m_W+m_{\widetilde{b}_1}$
\footnote{
We cannot easily set Br(${\widetilde{c_L}} \to e^+ d$) = 1.0 because 
${\widetilde{c_L}}$ can decay into $c\sz1$ via R-conserving interaction with 
a large branching fraction for a wide range of SUSY parameter 
space \cite{alt}. 
}. 
In this case it is difficult to discriminate the stop from one of 
the leptoquark $\widetilde{S}_{1/2}$ with the charge $Q=\frac{2}{3}$. 
Such a stop (or leptoquark) could be searched at Tevatron through the 
production mechanism, 
\begin{equation}
p{\overline{p}} \to \st {\overline{\st}} X \to e e X
\end{equation}
Recently $D0$ group has reported preliminary bounds on a leptoquark mass as 
$M_{LQ}$ $>$ 194GeV for Br($LQ \to e^{\pm} q$) = 1.0 \cite{D0}. 
As has been pointed out by Altarelli et al. \cite{alt} and 
Dreiner et al. \cite{drm}, 
detailed analyses at $D0$ and a joint analysis of the $D0$ and CDF 
dielectron data will be desirable for discovering or rejecting the 
stop (or leptoquark) with Br($\st \to e^+ d$) = 1.0.

On the other hand, 
if the $\widetilde{W}_1$ or $\sb$ is light, i.e.,  
$m_{\widetilde{W}_1}$ $\nle$ $m_{\widetilde{t}}-m_b$ or 
$m_{\sb}$ $\nle$ $m_{\st}-m_W$, 
Br($\widetilde{t} \rightarrow b\widetilde{W}_1$) or 
Br($\widetilde{t} \rightarrow \sb W$) can compete with 
Br($\widetilde{t} \rightarrow e^+d$). 
In this case, 
we should take into account of the processes 
\begin{eqnarray}
           ep         &\rightarrow&      b  \widetilde{W}_1 X, \\
           ep         &\rightarrow&      \sb  W X
\label{bc} 
\end{eqnarray}
It is expected that the detectable cross sections $\sigma 
\stackrel{>}{\sim}0.1$ pb 
for heavy stop with mass ${m_{\widetilde{t}}}\stackrel{<}{\sim} 250$ GeV for 
$e^+$ beams \cite{stopsg}. 
In our model the LSP, the lightest neutralino $\widetilde{Z}_1$, and $\sb$ 
possibly decay into $R$-even particles via only non-zero RB coupling 
$ \lambda'_{131} $.  
A typical decay chains will be  
\begin{eqnarray}
       ep \rightarrow b\widetilde{W}_1 X  &\rightarrow& 
       b(\ell\nu\widetilde{Z}_1) X  
       \rightarrow b(\ell\nu(bd\nu))X, \\
       ep \rightarrow \sb W X  &\rightarrow& (b\widetilde{Z}_1)W X  
       \rightarrow (b(bd\nu))(\ell\nu)X \\
        &\rightarrow& (\nu_e d)W X  
       \rightarrow (\nu_e d)(\ell\nu)X
\label{signature2}
\end{eqnarray}
   Thus, a possible typical signature of the stop production 
   $ep \rightarrow  b\widetilde{W}_1  X$ or $ep \rightarrow  \sb W  X$
   would be 
  2$b$-jets+jet+lepton+ ${\ooalign{\hfil/\hfil\crcr$P$}}_T$ or 
  jet+lepton+ ${\ooalign{\hfil/\hfil\crcr$P$}}_T$.   
  One of the signals to be detected at HERA is characterized by the high $P_T$ 
  spectrum of muons. 
  When such event excess could be observed, 
  the stop could clearly be distinguished from the leptoquark 
  $\widetilde{S}_{1/2}$.


Next we investigate the $Q^2$ and $M$ distributions of the expected events. 
In this analysis, we use the $Q_e^2$ and $M_e$ determined from 
energies and scattering angles of the final positron in combining the 
H1 and ZEUS data which are explicitly presented for the selected events 
refs.\cite{H1,ZEUS}. 

$Q^2_e$ and $M_e$ distributions of the expected number of events 
together with the experimental data are  
shown in Figs.2 and 3. 
Figure 2 corresponds to the case that only the lighter stop $\stl$ 
with mass $\mstl$ $=$ 210GeV is 
produced and the heavier one $\sth$ is decoupled ($\tht$ $=$ 0). 
While the event excess at high $Q^2_e$ is reproduced, 
the broad $M_e$ distribution cannot be well explained. 
It should be noted that the total decay width of the single stop 
is smaller than about 1.5GeV as mentioned in the event rate 
analysis. 
Consequently, the observed broad $M_e$ distribution would not be 
well explained by the single stop resonance scenario or by equivalent 
scharm ${\widetilde{c_L}}$ scenario. 

In Fig.3, on the other hand, we consider the case that two stops are 
almost degenarate in mass ($m_{\stl}$, $m_{\sth}$) $=$ (205GeV, 225GeV) 
with the finite mixing angle $\tht$ $=$ 0.95. 
In this case double resonance peaks could overlap each other and 
simulate the broad $M_e$ distribution. 
The event excess at high $Q^2_e$ of the data is also 
reproduced by this scenario. 
Difference between the one stop and the two stops scenario would be 
found in the very high $Q^2_e$ ($>$ 40,000GeV$^2$) region. 
For simplicity, we take Br($\st_{1,2} \to e^+ d$) = 1.0 in the calculation 
of Figs.2 and 3. 
It should be noted that the broader $M$ distribution can be obtained for 
Br($\st_{1,2} \to e^+ d$) $<$ 1.0 than for Br($\st_{1,2} \to e^+ d$) = 1.0. 

It has been pointed out that the ALEPH 4-jet events \cite{aleph} are possibly 
explained by the production mechanism 
$e^+e^-$ $\to$ ${\widetilde{e_L}}{\widetilde{e_R}}$ $\to$ 
$c{\overline{d}}{\overline{c}}d$ 
in the scharm scenario with light sleptons $m_{\sl}$ $\nle$ 60GeV 
\cite{carena}. 
We have discussed the 4-jet events in the stop scenario \cite{koba}. 
In this case possible mechanism is 
$e^+e^-$ $\to$ ${\widetilde{\nu_e}}{\overline{\widetilde{\nu_e}}}$ $\to$ 
$b{\overline{d}}{\overline{b}}d$. 
Recent analysis of the 4-jet events seems to reveal 
no $b$-quark enhancements and pair production of equal mass particles 
will be disfavoured. 
If ths is the case, our stop scenario may have no relation to 
the ALEPH 4-jet events.


We have investigated a possible scenario to explain 
an event excess in the reaction $e^+p \to e^+ X$ with large $x$ and high 
$Q^2$ observed 
by the H1 and ZEUS Collabolation at HERA by the resonance production of 
the stops with an 
$R$-parity breaking 
interaction in the framework of the MSSM. 
The HERA events are characterized by not only high $Q^2$ but also the broad 
$M$ distribution. 
It can be simulated by our specific scenario if we 
consider almost degenerate two mass eigenstates $\widetilde{t_1}$ and 
$\widetilde{t_2}$ 
of the stops. 

Our scenario would be confirmed or rejected at 
HERA or Tevatron through the search for 
the high $P_T$ muon events (HERA) or 
the high $P_T$ $ee$ events (Tevatron). 
Moreover, in our favourable scenario, 
$m_{\stl,\sth}$ $\simeq$ 210GeV, 
one of sbottom ${\widetilde{b_L}}$ could be light as 
120GeV $\sim$ 150GeV and it can decay into $\nu_e d$ via 
the RB interactions proportional to $\lambda'_{131}$. 
It will be a good target of LEP2 or future linear colliders. 
Detailed studies of the sbottom production at $e^+e^-$ colliders 
in our specific scenario will be reported soon. 
Needless to say, for our purpose it would be highly desirable to carry out 
the high luminosity run at HERA for making sure the broad $M$ distribution.

\begin{flushleft}
{\Large{\bf Acknowledgements}}
\end{flushleft}
One of the present authors (T. Kon) was supported in part by 
the Grant-in-Aid for Scientific Research from the Ministry of Education, 
Science and Culture of Japan, No. 08640388.



\vfill\eject

{\Large{\bf Figure Captions}}
\begin{description}

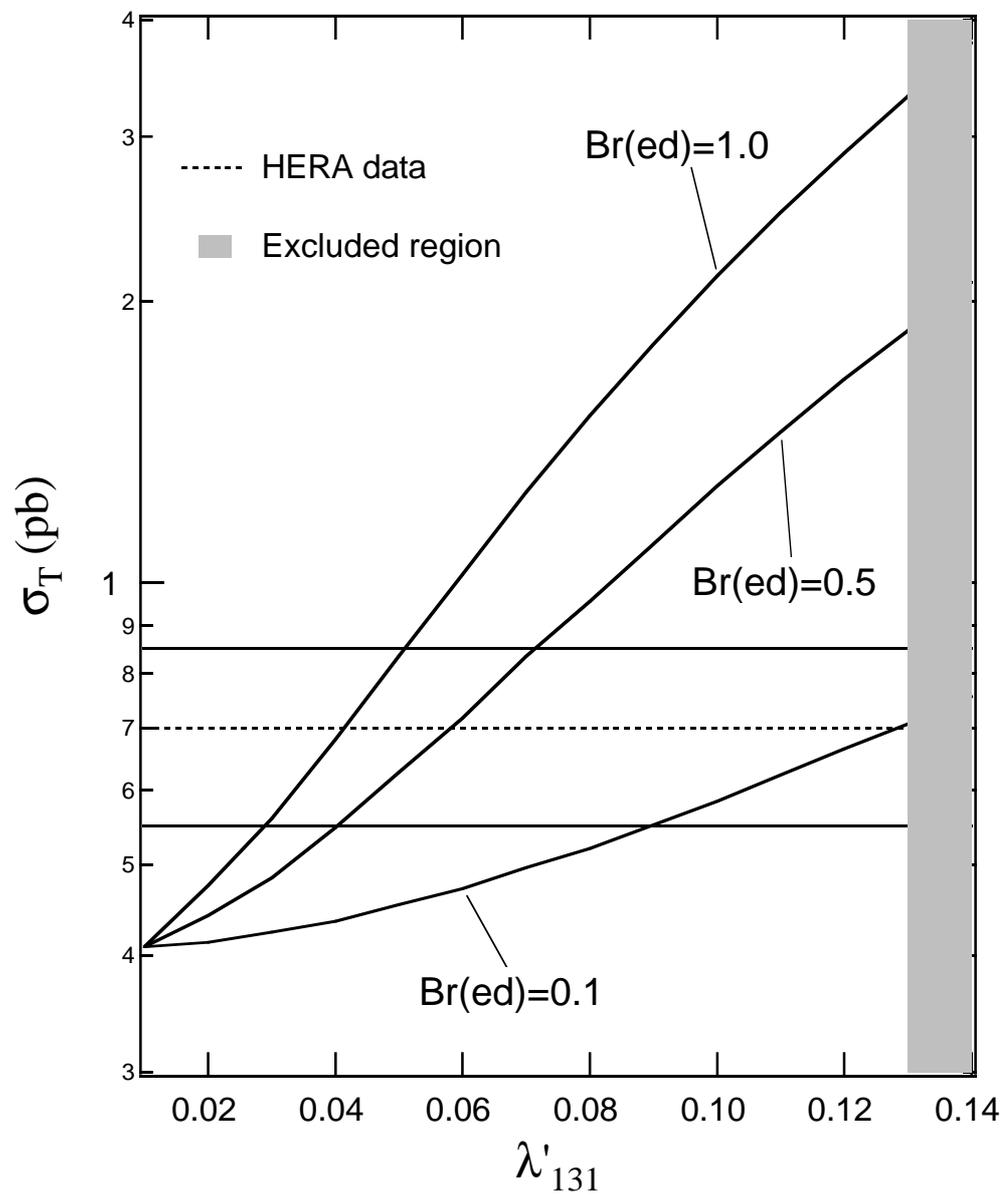
\item[{\bf Figure 1:}]
$\lambda'_{131}$ dependence of the total cross section  
$\sigma (e^+ p \to \stl X \to e^+ q X)$. 
We take 
$\mstl$ $=$ 210GeV, $\tht$ $=$ 0 and $Q^2$ $>$ 15,000GeV$^2$. 
The horizontal lines correspond to the experimental value at HERA, 
$\sigma_{obs}$ $=$ (0.71 $\pm$ 0.15) pb. 

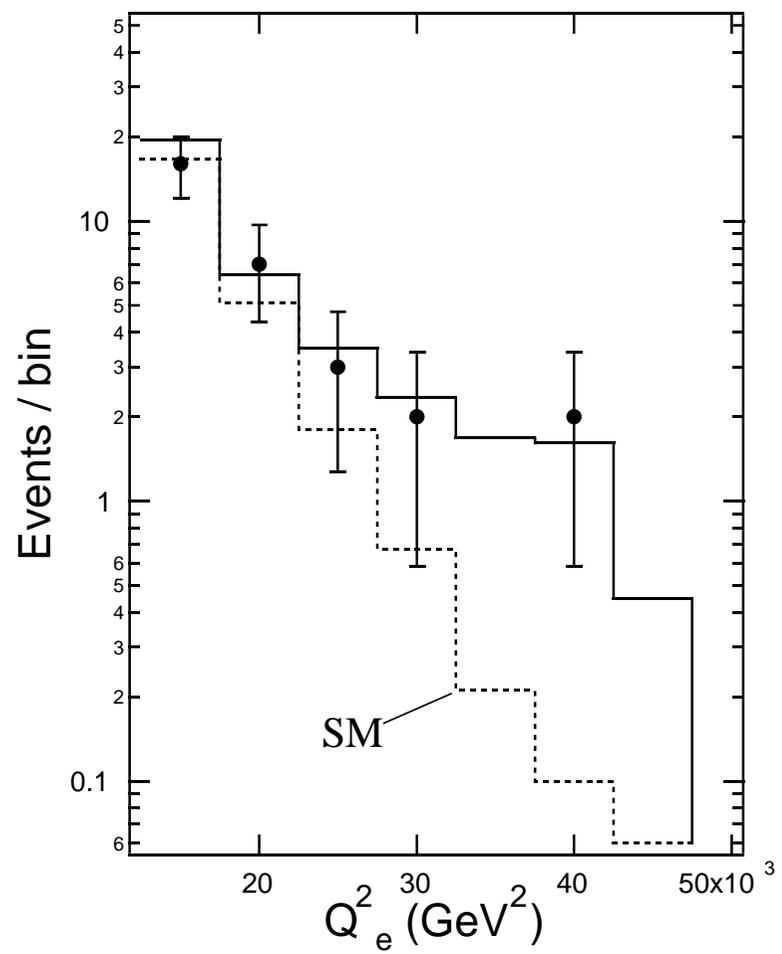
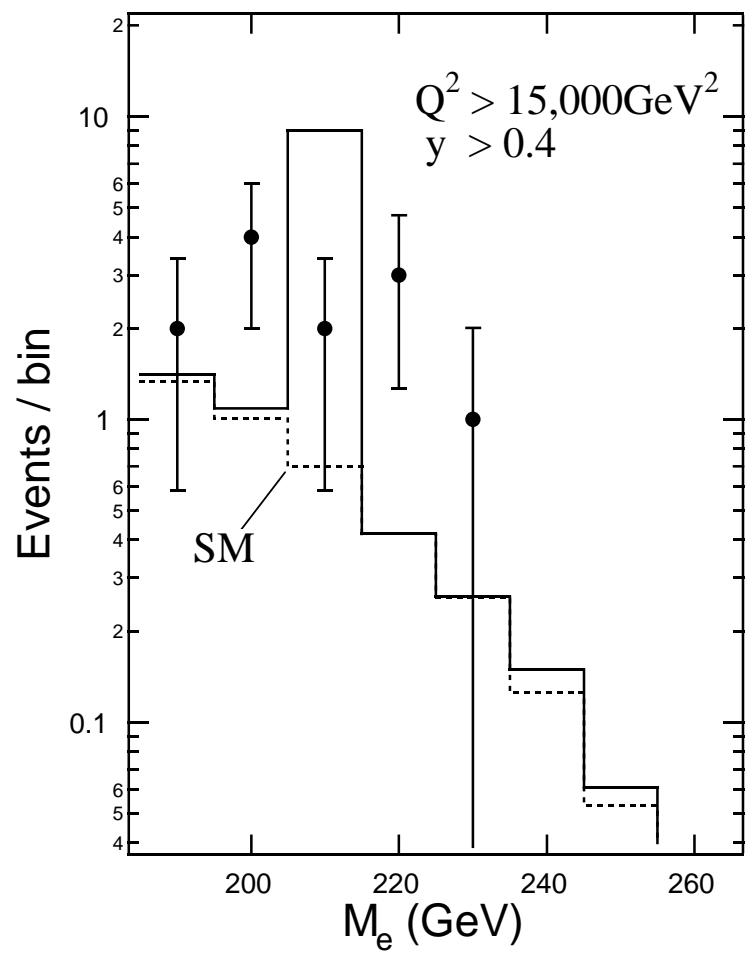
\item[{\bf Figure 2:}]
$Q^2_e$ and $M_e$ distribution of the expected number of events 
together with the experimental data.
We take $\mstl$ $=$ 210GeV, $\tht$ $=$ 0, $\lambda'_{131}$ $=$ 0.04 
and integrated luminosity $L=34$pb$^{-1}$. 
For $M_e$ distribution $Q^2$ $>$ 15,000GeV$^2$ and $y_e$ $>$ 0.4 are 
adopted. 
Dashed line corresponds to the SM expectation.

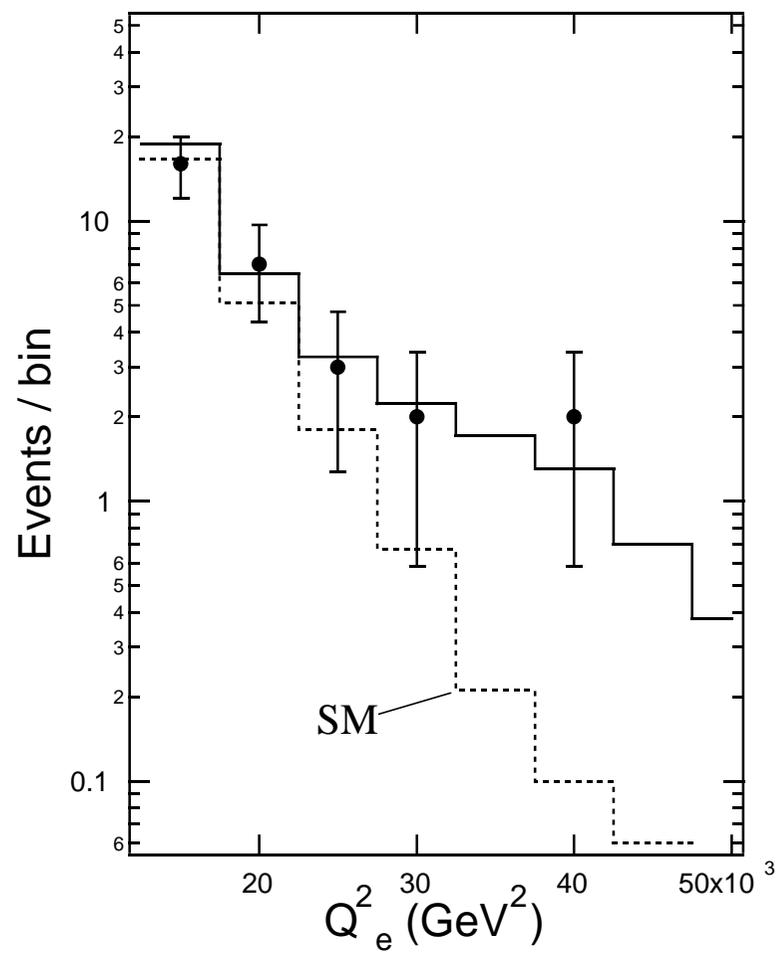
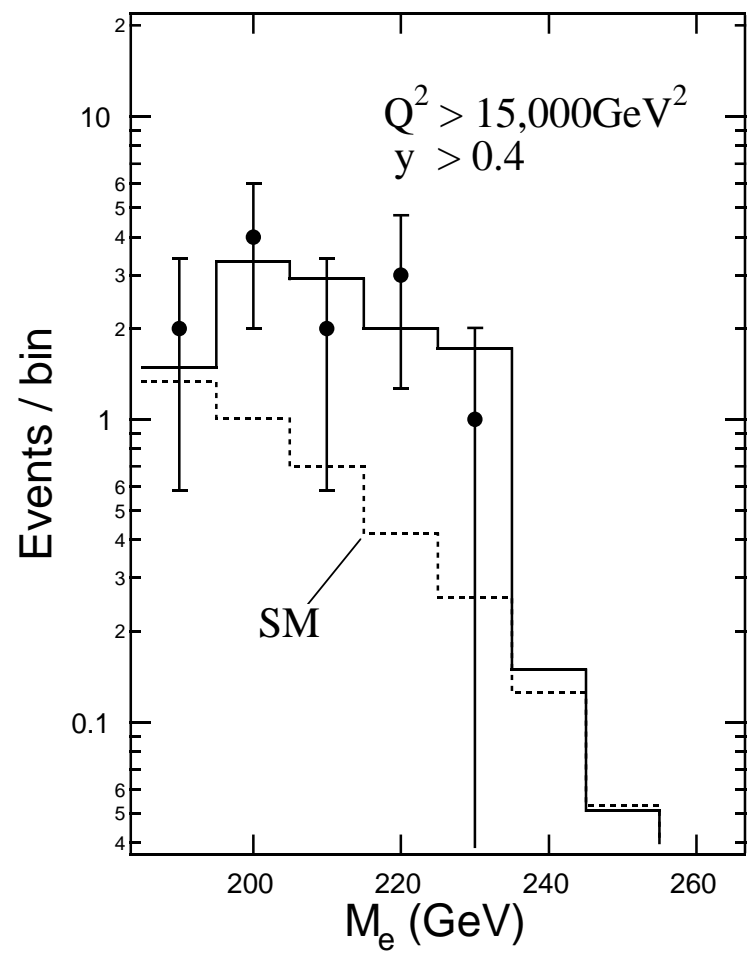
\item[{\bf Figure 3:}]
$Q^2_e$ and $M_e$ distribution of the expected number of events 
together with the experimental data.
We take $\mstl$ $=$ 205GeV, $\msth$ $=$ 225GeV, $\tht$ $=$ 0.95, 
$\lambda'_{131}$ $=$ 0.045 
and integrated luminosity $L=34$pb$^{-1}$. 
For $M_e$ distribution $Q^2$ $>$ 15,000GeV$^2$ and $y_e$ $>$ 0.4 are 
adopted. 
Dashed line corresponds to the SM expectation. 
\end{description}

\vfill\eject

\end{document}